\documentclass[conference,10pt]{IEEEtran}
\usepackage{graphicx}
\usepackage{epstopdf}
\usepackage{graphicx}
\usepackage{amsmath}
\usepackage{amsfonts}
\usepackage{amssymb, color, bm}

\usepackage{algorithmicx}
\usepackage[noend]{algpseudocode}
\usepackage{algorithm,setspace}
\usepackage[T1]{fontenc}% optional T1 font encoding

\usepackage{multicol}
\usepackage{lipsum}
\usepackage{mwe}

\DeclareMathOperator{\E}{\mathbb{E}}

\usepackage[utf8]{inputenc}
\allowdisplaybreaks
\usepackage{cite}
\DeclareMathOperator{\R}{\mathbb{R}}
\DeclareMathOperator{\I}{\mathbb{I}}

\usepackage{algorithm}
\usepackage[noend]{algpseudocode}

\usepackage{lineno}
\title{Fast Power Control Adaptation via Meta-Learning for Random Edge Graph Neural Networks}
\author{
    \IEEEauthorblockN{Ivana Nikoloska and Osvaldo Simeone}
    \IEEEauthorblockA{KCLIP, CTR, Dept. of Engineering, King's College London
    \\\{ivana.nikoloska, osvaldo.simeone\}@kcl.ac.uk}
    }

\begin{document}

\maketitle
\let\thefootnote\relax\footnote{This work was supported by the European Research Council (ERC) under the European Union’s Horizon 2020 Research and Innovation Program (Grant
Agreement No. 725731).}

\begin{abstract}
Power control in decentralized wireless networks poses a complex stochastic optimization problem when formulated as the maximization of the average sum rate for arbitrary interference graphs. Recent work has introduced data-driven design methods that leverage graph neural network (GNN) to efficiently parametrize the power control policy mapping channel state information (CSI) to the power vector. The specific GNN architecture, known as random edge GNN (REGNN), defines a non-linear graph convolutional architecture whose spatial weights are tied to the channel coefficients, enabling a direct adaption to channel conditions. This paper studies the higher-level problem of enabling fast adaption of the power control policy to time-varying topologies. To this end, we apply first-order meta-learning on data from multiple topologies with the aim of optimizing for a few-shot adaptation to new network configurations. 
\end{abstract}


\section{Introduction}
Power and bandwidth are fundamental resources in communication systems, playing a key role in determining the effective capacity of a wireless channel. The optimal allocation of these resources under time-varying channel characteristics and user demands is essential to efficiently scale wireless systems. As a notable example, power allocation in a wireless ad-hoc network is crucial to mitigate multi-user interference, which is often the performance bottleneck \cite{hossain2014evolution}, \cite{hong2014applications}. However, solving the radio resource management problem in its most general form is NP-hard, implying that, as the network becomes denser, it becomes more challenging to derive an optimal solution \cite{mollanoori2013uplink}.

To deal with these challenges, many approaches have been proposed in the literature. These range from classical optimization techniques \cite{lei2015joint} to information and game theory \cite{yang2017mean,riaz2018power}, and tackle various radio resource management subproblems. Recent advances in machine learning offer a promising framework in which to develop solutions in the presence of model and/or algorithmic deficits \cite{simeone2018very}.  However, the performance of the trained models generally depend on how representative the training data are for the channel conditions encountered at deployment time. As a result, when conditions in the network change, these rigid models often are no longer useful \cite{nair2019covariate}, \cite{quinonero2009dataset}.

This problem was successfully addressed by the data-driven methodology introduced in \cite{eisen2020optimal}, and also studied in \cite{naderializadeh2020wireless,chowdhury2020unfolding}. In it, the power control policy mapping channel state information (CSI) and power vector is parametrized by a graph neural network (GNN). The GNN encodes information about the network topology in its underlying graph, and it applies spatial weights that are tied to the channel realizations. The design problem consists of training the shared temporal weights of the graph filters applied by the GNN. By tying the spatial weights to the CSI, the solution -- which is referred to as random edge GNN (REGNN) -- automatically adapts to varying CSI conditions.

In this paper, we focus on the higher-level problem of facilitating adaptation to time-varying topologies. To this end, as illustrated in Fig.~\ref{sys_mod}, we assume that the topology of the network varies across periods of operation of the system, with each period being characterized by time-varying channel conditions. As such, the operation within each channel period is well reflected by the model studied in \cite{eisen2020optimal,naderializadeh2020wireless}, and we adopt an REGNN architecture for within-period adaptation. We assume that the network designer is given limited CSI data at the beginning of each period that can be used to adapt the temporal filter to the changed topology. In order to facilitate fast adaptation -- in terms of data and iteration requirements -- we integrate meta-learning with REGNN training.

%Specifically, we consider first-order meta-learning methods in order to enable the models to quickly adapt to dynamic networks in which both the number of the nodes, as well as the position of the nodes changes over time. To this end, we adopt the Random Edge Graph Neural Network  in order to parameterize the power allocation policy, 
Meta-learning leverages CSI data from a number of previous periods to optimize an inductive bias that facilitates fast adaptation on a new topology to be encountered in a future period. We specifically adopt first-order meta-learning methods \cite{finn2017model}, \cite{nichol2018firstorder} that encode the inductive bias in the initialization of the adaptation procedure within each period. While GNNs are known to be robust to changes in the topology, the proposed integration of meta-learning and REGNNs is shown to offer significant improvements in terms of sample and iteration efficiency.     

Previous work on meta-learning for wireless system includes \cite{park2020learning}, \cite{jiang2019mind}, \cite{yuan2020transfer}, which have focused on the problem of optimizing demodulation, decoding, and beamforming, respectively. The application to GNN-based power control is presented here for the first time.

The rest of the paper is organized as follows. The considered model and problem are presented in Section~\ref{sec_mod_prob}, and REGNNs are reviewed in Section~\ref{sec_regnn}. The adopted meta-learning methods are given in Section~\ref{sec_meta}, and are evaluated in Section~\ref{sec_num}. Section~\ref{sec_con} concludes the paper.

\section{Model and Problem}\label{sec_mod_prob}
\begin{figure*}[tbp]
\centering
\includegraphics[width=0.86\linewidth]{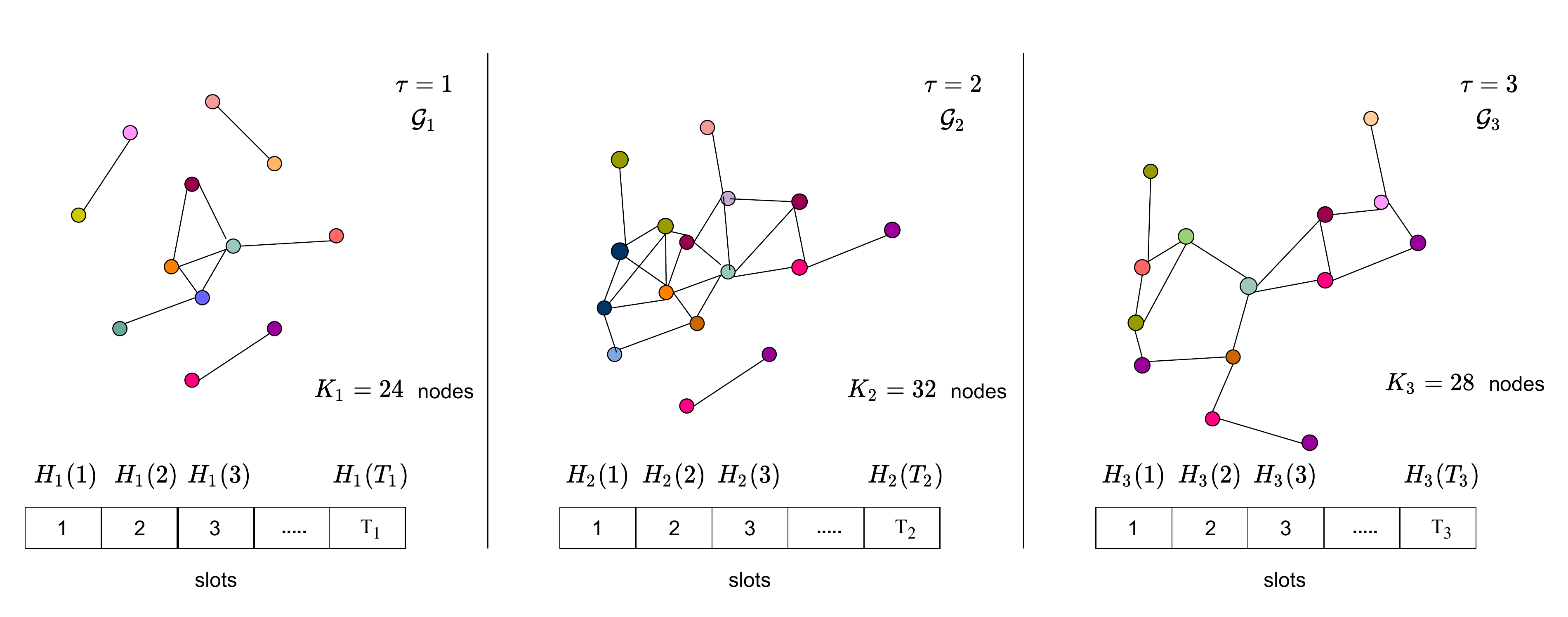}
\vspace*{-3mm}
\caption{Interference graph $\mathcal{G}_\tau$ over periods $\tau = 1$, $\tau = 2$, and $\tau = 3$. Each vertex represents a communication link, and an edge is included between interfeering links. }
\label{sys_mod}
\end{figure*}

As illustrated in Fig.~\ref{sys_mod}, we consider a wireless network running over periods $\tau=1,...,\mathcal{T}$, with topology possibly changing at each period $\tau$. During period $\tau$, the network is comprised of $K_\tau$ communication links. Transmissions on the $K_\tau$ links are assumed to occur at the same time using the same frequency band.  The resulting interference graph $\mathcal{G}_\tau = (\mathcal{K}_\tau, \mathcal{E}_\tau)$ includes an edge $(k, j) \in \mathcal{E}_\tau$ for any pair of links $k, j \in \mathcal{K}_\tau$ with $i \neq j$ whose transmissions interfere with one another. Both the number of links $K_\tau = |\mathcal{K}_\tau|$ and the graph $\mathcal{G}_\tau$ generally vary across periods $\tau$. We denote by $\mathcal{N}^k_\tau \subseteq \mathcal{K}_\tau$ the subset of links that interfere with link $k$ at period $\tau$.

Each period contains $T_\tau$ time slots, indexed by $t = 1,...,T_\tau$. In time slot $t$ of period $\tau$, the channel between the transmitter of link $k$ and its intended receiver is denoted by $h^{k,k}_\tau (t)$, while $h^{j,k}_\tau (t)$ denotes the channel between transmitter of link $j$ and receiver of link $k$ with $j \in \mathcal{N}^k_\tau$. We have that $h^{j,k}_\tau (t) = 0$ for $j \notin \mathcal{N}^k_\tau$. The channels for slot $t$ in period $\tau$ are arranged in the channel matrix $H_\tau (t) \in \R^{K_\tau \times K_\tau}$, with the $(j,k)$ entry given by $\left[ H_\tau(t) \right]_{j,k} = h^{j,k}_\tau (t)$. 
Channel states vary across time slots, and the marginal distribution of matrix $H_\tau(t)$ for all $t = 1,...,T_\tau$ is constant and denoted by $\mathcal{P}_\tau (H_\tau)$. The distribution $\mathcal{P}_\tau (H_\tau)$ generally changes across periods $\tau$ and it is a priori unknown to the network.

To manage the inter-link interference, it is useful to adjust the transmit powers such that a global network-wide objective function is optimized. For each channel realization $H_\tau (t)$, we denote the vector of power allocation variables by $p_\tau (t) \in \R^{K_\tau}$, whose $k$-th component, $p^k_\tau (t)$, represents the transmit power of transmitter $k$ at time slot $t$ of period $\tau$. The resulting achievable rate for link $k$ is given by
\begin{align}\label{rate_k}
    c^k(H_\tau (t), p_\tau (t)) = \log \left(1 + \frac{|h^{k,k}_\tau (t)|^2 p^k_\tau (t)}{\sigma^2 + \sum_{j \in \mathcal{N}_\tau^k} |h^{j,k}_\tau(t)|^2 p^j_\tau(t)}\right),
\end{align}
where $\sigma^2$ denotes the noise power. 

The goal of the system is to determine a power allocation policy $\textrm{p}_\tau(\cdot)$ in each period $\tau$ that maps the channel matrix $H_\tau(t)$ to a power allocation vector $p_\tau (t)$ by maximizing the average achievable sum-rate as
\begin{align}\label{opt}
    &\underset{\textrm{p}_\tau(\cdot)}{\text{max}} \,\,\,\, \sum_{k = 1}^{K} \E_{H_\tau \sim \mathcal{P} (H_\tau)} \Big[c^k(H_\tau, p_\tau(H_\tau))\Big] \nonumber\\
    & \text{s.t.} \,\,\,\,  0 \leq \textrm{p}^k_\tau(\cdot) \leq P^k_{\textrm{max}}, \,\,\,\, \text{for} \,\,\,\, k = 1,...,K,
\end{align}
where $P^k_{\textrm{max}}$ denotes the power constraint of link $k$. Note that, the problem is defined separately for each period $\tau$. The distribution $\mathcal{P} (H_\tau)$ is unknown, and the designer has access only to channel realizations $\{H_\tau (1), ..., H_\tau (T_\tau)\}$ over $T_\tau$ time slots. Accordingly, problem \eqref{opt} is approximated as 
\begin{align}\label{opt_approx}
    &\underset{\textrm{p}_\tau(\cdot)}{\text{max}} \,\,\,\, \sum_{k = 1}^{K} \sum_{t = 1}^{T_\tau} c^k(H_\tau (t), p_\tau(H_\tau(t))) \nonumber\\
    & \text{s.t.} \,\,\,\,  0 \leq \textrm{p}^k_\tau(\cdot) \leq P^k_{\textrm{max}}, \,\,\,\, \text{for} \,\,\,\, k = 1,...,K.
\end{align}

\section{Power Allocation by Training REGNN}\label{sec_regnn}
In this section, we review the solution proposed in \cite{eisen2020optimal}, which tackles problem \eqref{opt_approx}
separately for each period $\tau$. Accordingly, as we will see, the method requires a sufficiently large data set $\mathcal{D}_\tau$ to be available for each period $\tau$. The method parameterises the function $\textrm{p}_\tau(\cdot)$ by a REGNN. Specifically, one sets $p_\tau(H_\tau)  = \textrm{f} (H_\tau \, | \, \phi_\tau)$, where $\phi_\tau \in \R^M$ is a vector of trainable parameters that defines the operation of the REGNN $\textrm{f} (H_\tau \, | \, \phi_\tau)$. To simplify notation, in this section, we drop the index $\tau$, which is fixed.

The  REGNN $\textrm{f} (H \, | \, \phi)$ alternates linear graph filters and pointwise non-linearities. It generalises the operation of convolutional neural networks (CNNs) \cite{krizhevsky2012imagenet} by implementing a convolution on graph-structured data. To describe the REGNN, we first define the graph convolution operation 
\begin{align}\label{regnn}
    \phi \ast_H x =  \sum_{m=1}^M \phi_{m} H^m x,
\end{align}
where $\phi = [\phi_1, ... , \phi_M]^T$ is a vector of filter taps with $\phi_m \in \R$; $x \in \R^{K}$ denotes the input signal; and $H^m$ denotes the $m$-th power of the channel matrix $H \in \R^{K \times K}$. The REGNN consists of a layered architecture where the output of layer $l-1$ is fed as an input to layer $l$. Specifically, the output of the $l$-th intermediate layer is given as
\begin{align}\label{regnn1}
    z_{l+1} = \sigma \left[ \phi \ast_H z_{l} \right],
\end{align}
where $\sigma [\cdot]$ denotes a pointwise non-linearity. 
The REGNN is defined by recursive application of \eqref{regnn} for $L$ layers, with $z_0 = x$. The input signal is set to an all-one vector \cite{naderializadeh2020wireless}, and it may more generally include a variable describing the state of the link \cite{eisen2020optimal}. The transmit power is found as the output of the REGNN as
\begin{align}\label{regnn2}
    &\textrm{f} (H \, | \, \Phi) = P_\text{\textrm{max}} \nonumber\\
    &\times \sigma \left[ \sum_{m=1}^M \phi_{L,m} H^m \left( ... \left(\sigma \left[ \sum_{m=1}^M \phi_{1,m} H^m x \right]\right) ...\right) \right],
\end{align}
with $P_\text{\textrm{max}}$ being a diagonal matrix with its $k$-th element on the main diagonal being given by $P^k_{\textrm{max}}$, and $\phi_{l,m}, \, m = 1,...,M$ denoting the parameters of layer $l$. The REGNN \eqref{regnn2} can be described as applying a message-passing procedure on the interference graph. Messages exchanged at each layer are weighted by the relevant entries of the powers $H^m$ of the channel matrix. Therefore, due to its dependence on the random fading channels the GNN is characterized by "random edges" \cite{eisen2020optimal}.

Given a set of channel realizations $\{H (1), ..., H (T)\}$ for a given period, training of the model parameters $\Phi = [\phi_{l,m}]_{l=1,...,L; \, m=1,...,M} \in \R^{L \times M}$ is done by tackling problem
\begin{align}\label{opt_phi}
    &\underset{\Phi}{\text{max}} \,\,\,\, \sum_{k = 1}^{K} \sum_{t = 1}^{T}  c^k(H (t), \textrm{f} (H (t) \, | \, \Phi)),
\end{align}
via stochastic gradient descent (SGD). It is noted that, by incorporating the channel matrix in the structure of the REGNN-based power control policy $\textrm{p}(\cdot)$, the method proposed in \cite{eisen2020optimal} automatically adapts to the different per-slot channel realizations. 

%\begin{figure}[tbp]
%\centering
%\includegraphics[width=0.75\linewidth]{figures/ML_obj.pdf}
%\caption{The objective of FOMAML and REPTILE is to identify the initial shared parameters, $\Theta$, such that when initiating an optimization from this point, a few SGD updates produce the task-specific parameters $\Phi_1$, $\Phi_2$ and $\Phi_3$ that yield high communication rate using a small number of data samples. }
%\label{ML_obj}
%\end{figure}

\section{Fast Per-Period Adaptation via Meta-Learning}\label{sec_meta}
In this section, we introduce the proposed meta-learning solution. The main goal is to improve the data efficiency of the solution reviewed in the previous section by enabling explicit  adaptation of the power control policy $\textrm{p}_\tau(\cdot)$ for each period $\tau$, and hence across the changing topologies (see Fig.~\ref{sys_mod}). In this regard, it is noted that the approach reviewed in the previous section already has some robustness properties to changes of the network topology \cite{eisen2020optimal}. However, in practice, as we will see, better results can be obtained by explicitly adapting the  power control policy $\textrm{p}_\tau(\cdot)$ to the current topology defined by the interference graph $\mathcal{G}_\tau$.
%as this approach can yield good performance even when the number of nodes in the network is not fixed and permutation invariance does not hold.

In order to enable (offline) meta-learning, we leverage data from $\mathcal{T}^{\text{meta}}$ periods, which we denote by $\mathcal{D} = \{\mathcal{D}_\tau\}_{\tau = 1,...,\mathcal{T}^{\text{meta}}}$, with $\{\mathcal{D}_\tau\} = \{H_\tau (1), ..., H_\tau (T_\tau)\}$.
Following standard practice in meta-learning, each meta-training data set $\mathcal{D}_\tau$ is split into training data $\mathcal{D}^{\text{tr}}_\tau$ and testing data $\mathcal{D}^{\text{te}}_\tau$ \cite{finn2017model}, \cite{simeone2020learning}, and we write $t \in \mathcal{D}^{\text{tr}}_\tau$ and $t \in \mathcal{D}^{\text{te}}_\tau$ to denote the indices of the slots assigned to each set. At test time, during deployment, the network observes a new topology $\mathcal{G}_{\tau_\text{test}}$ for which it has access to a data set $\mathcal{D}_{\tau_\text{test}}$, which is generally small.

\subsection{FOMAML}
In model agnostic meta learning (MAML) the key idea is to identify an initialization vector $\Phi_0 \in \R^M$ for the model parameters $\Phi_\tau \in \R^M$ that enables fast adaptation in each period $\tau$. This is in the sense that only a few SGD updates for problem \eqref{opt_phi}, based on limited, per-period data, produce period-specific parameters $\Phi_\tau$ that yield high communication rates. For example, with a single SGD step and a full mini-batch update, the model parameters in period $\tau$ are updated as
\begin{align}\label{locM}
    \Phi_\tau = \Phi_0 + \eta \nabla_{\Phi_0} \left( \sum_{t \in \mathcal{D}^{\text{tr}}_\tau} \sum_{k = 1}^{K} c^k(H_\tau(t), \textrm{f} (H_\tau(t) \, | \, \Phi_0))\right),
\end{align}
starting from initialization $\Phi_0$ where $\eta > 0$ denotes the learning rate. 

The shared initialization vector $\Phi_0$ is identified by maximizing the rate achieved across all meta-training data sets upon adaptation steps of the form in \eqref{locM}, or, more generally, with multiple SGD updates. Specifically, the objective is given as
\begin{align}\label{meta_opt_phi}
    &\underset{\Phi_0}{\text{max}} \,\,\,\, \sum_{k = 1}^{K_\tau} \sum_{\tau = 1}^{\mathcal{T}^{\text{meta}}} \sum_{t \in \mathcal{D}^{\text{te}}_\tau} c^k(H (t), \textrm{f} (H (t) \, | \, \Phi_\tau)) \nonumber\\
    &= \sum_{k = 1}^{K_\tau} \sum_{\tau = 1}^{\mathcal{T}^{\text{meta}}} \sum_{t \in \mathcal{D}^{\text{te}}_\tau}  c^k \Bigg(H (t), \textrm{f} \Bigg(H (t) \, | \, \Phi_0 \nonumber\\
    &+\eta \nabla_{\Phi_0} \Bigg( \sum_{k = 1}^{K_\tau} \sum_{t \in \mathcal{D}^{\text{tr}}_\tau} c^k(H_\tau(t), \textrm{f} (H_\tau(t) \, | \, \Phi_0))\Bigg)\Bigg)\Bigg),
\end{align}
where in the second equality we have used the single SGD update for simplicity of notation.
The generalization of \eqref{meta_opt_phi} to an arbitrary number of SGD updates is direct. The key idea is that an initial $\Phi_0$ maximizing \eqref{locM} should also enable fast adaptation at test time.
Tackling \eqref{meta_opt_phi} via SGD requires updating the shared parameters as
\begin{align}\label{locMsecond}
    \Phi_0 &\leftarrow \Phi_0 \nonumber\\
    &- \epsilon \left( \I - \nabla^2_{\Phi_0} \sum_{k = 1}^{K_\tau} \sum_{\tau = 1}^{\mathcal{T}^{\text{meta}}} \sum_{t \in \mathcal{D}^{\text{te}}_\tau} c^k(H_\tau(t), \textrm{f} (H_\tau(t) \, | \, \Phi_0)) \right) \nonumber \\
    &\times \left( \nabla_{\Phi_\tau} \sum_{k = 1}^{K_\tau} \sum_{\tau = 1}^{\mathcal{T}^{\text{meta}}} \sum_{t \in \mathcal{D}^{\text{te}}_\tau} c^k(H_\tau(t), \textrm{f} (H_\tau(t) \, | \, \Phi_\tau)) \right),
\end{align}
where $\I$ denotes the identity matrix and $\epsilon > 0$ denotes the learning rate. First-order MAML (FOMAML) ignores the Hessian terms in the updates of the shared parameters in \eqref{locMsecond}, obtaining the update
\begin{align}
    \Phi_0 &\leftarrow \Phi_0 \nonumber\\
    &+ \epsilon \nabla_{\Phi_\tau} \left( \sum_{k = 1}^{K_\tau} \sum_{\tau = 1}^{\mathcal{T}^{\text{meta}}} \sum_{t \in \mathcal{D}^{\text{te}}_\tau}  c^k(H_\tau(t), \textrm{f} (H_\tau(t) \, | \, \Phi_\tau)) \right).
\end{align}
%Note that, the task-specific parameters are updated using the training portion of the meta-training data set $\{\mathcal{D}^{\text{tr}}_\tau\}$, whilst the task-specific parameters are updated using the testing portion $\{\mathcal{D}^{\text{te}}_\tau\}$ to avoid over-fitting.

\subsection{REPTILE}
As a first-order meta-learning algorithm, REPTILE also learns an initialization for the parameters $\Phi_\tau$ by circumventing computation of higher-order derivatives. The shared parameters are specifically updated as \cite{nichol2018firstorder}
\begin{align}
    \Phi_0 \leftarrow (1-\epsilon) \Phi_0 - \epsilon \left( \frac{1}{\mathcal{T}^{\text{meta}}}\sum_{\tau = 1}^{\mathcal{T}^{\text{meta}}} \Phi_\tau \right),
\end{align}
where $\epsilon > 0$ denotes the learning rate.

\section{Numerical Evaluation}\label{sec_num}
In this section, we provide numerical results on meta-learning procedures for power allocation in distributed wireless networks. Code will be made available at \cite{git}.

\subsection{Network and Channel Model}
A random geometric graph in two dimensions comprised of $2 K_\tau$ nodes is drawn in each period $\tau$ by dropping transmitter $k$ uniformly at random at location $\textrm{Tx}_{\tau,k} \in \left[-K_\tau, K_\tau \right]^2$, with its paired receiver $r^k_\tau$ at location $\textrm{Rx}_{\tau,k} \in \left[\textrm{Tx}_{\tau,k}-\frac{K_\tau}{4}, \textrm{Tx}_{\tau,k} + \frac{K_\tau}{4}\right]^2$. Given the geometric placement, the fading channel state between transmitter $k$ and receiver $j$ is given by $h^{k,j}_\tau (t) = h^{k,j}_{\tau, p} (t) h^{k,j}_{\tau, f} (t)$, where the subscript $p$ denotes the path-loss gain, and the subscript $f$ denotes the fast-fading component, which depend on the time slot $t$. The constant path-loss gain can be found as $h^{k,j}_{\tau, p} = ||\textrm{Tx}_{\tau,k} - \textrm{Rx}_{\tau,k} ||^{-\gamma}$, where the path-loss exponent is set to $\gamma = 2.2$. The fast fading component $h^{k,j}_{\tau, f} (t)$ is random, and is drawn i.i.d. over indices $t$ and $\tau$ according to $|h^{k,j}_{\tau, f} (t)| \sim \text{Rayleigh} \,\, (\alpha)$, where we set $\alpha = 1$. Thereby, at each time slot $t$, fading conditions change, and the instantaneous channel information is used by the model to generate the optimal power allocation. The noise power $\sigma^2$ is set to $\sigma^2 = -70$dBm, and the maximum transmit power $P^k_{\text{max}}$ is set to $P^k_{\text{max}} = -35$dBm for all devices.

\subsection{Model Architecture and Hyperparameters}
As in \cite{naderializadeh2020wireless}, we consider a REGNN comprised of $L = 2$ hidden layers, each containing a filter of size $M = 4$. The non-linearity $\sigma [\cdot]$ in \eqref{regnn1} and \eqref{regnn2} is a Rectified Linear Unit (ReLU), given by $\sigma(x) = \text{max}(0,x)$, except for the output layer where we use a sigmoid. In all experiments we set the input signal to an all-one vector.

\subsection{Data sets}
We study separately the case where the number of the nodes in the network $K_\tau$ is fixed, but the topology changes across periods, as well as the case when the number of nodes in the network is also time-varying.

\subsubsection{Fixed Network Size}
In the first scenario, for a fixed number of links $K_\tau = 10$, each meta-training data set $\mathcal{D}_\tau$ corresponds to the realization of the random drop of the transmitter-receiver pairs at period $\tau$. Each drop is then run for $T_\tau = 400$ slots, whereby the fading coefficients are sampled i.i.d. at each slot. The training and testing portions of the data set $\mathcal{D}_\tau$ contain $200$ slots each.
\subsubsection{Dynamic Network Size}
In the second scenario, the size of the network is chosen uniformly at random as $K_\tau \sim \text{Uniform} \,\, [4, 20]$. Each meta-training data set $\mathcal{D}_\tau$ corresponds to a realization of the network size and to a random drop of the transmitter-receiver pairs. The fading coefficients are sampled randomly at each slot as discussed above.

In both scenarios, we set the number of meta-training periods $\mathcal{T}^{\text{meta}}$ to $\mathcal{T}^{\text{meta}} = 50$. The number of samples in the data sets $\mathcal{D}^{\text{tr}}_\tau$, and $\mathcal{D}^{\text{te}}_\tau$ are set to $|\mathcal{D}^{\text{tr}}_\tau| = 200$, and $|\mathcal{D}^{\text{te}}_\tau| = 200$, respectively, for $\tau = 1,...,\mathcal{T}^{\text{meta}}$.

\subsection{Results}
\begin{figure}[tbp]
\centering
\includegraphics[width=0.85\linewidth]{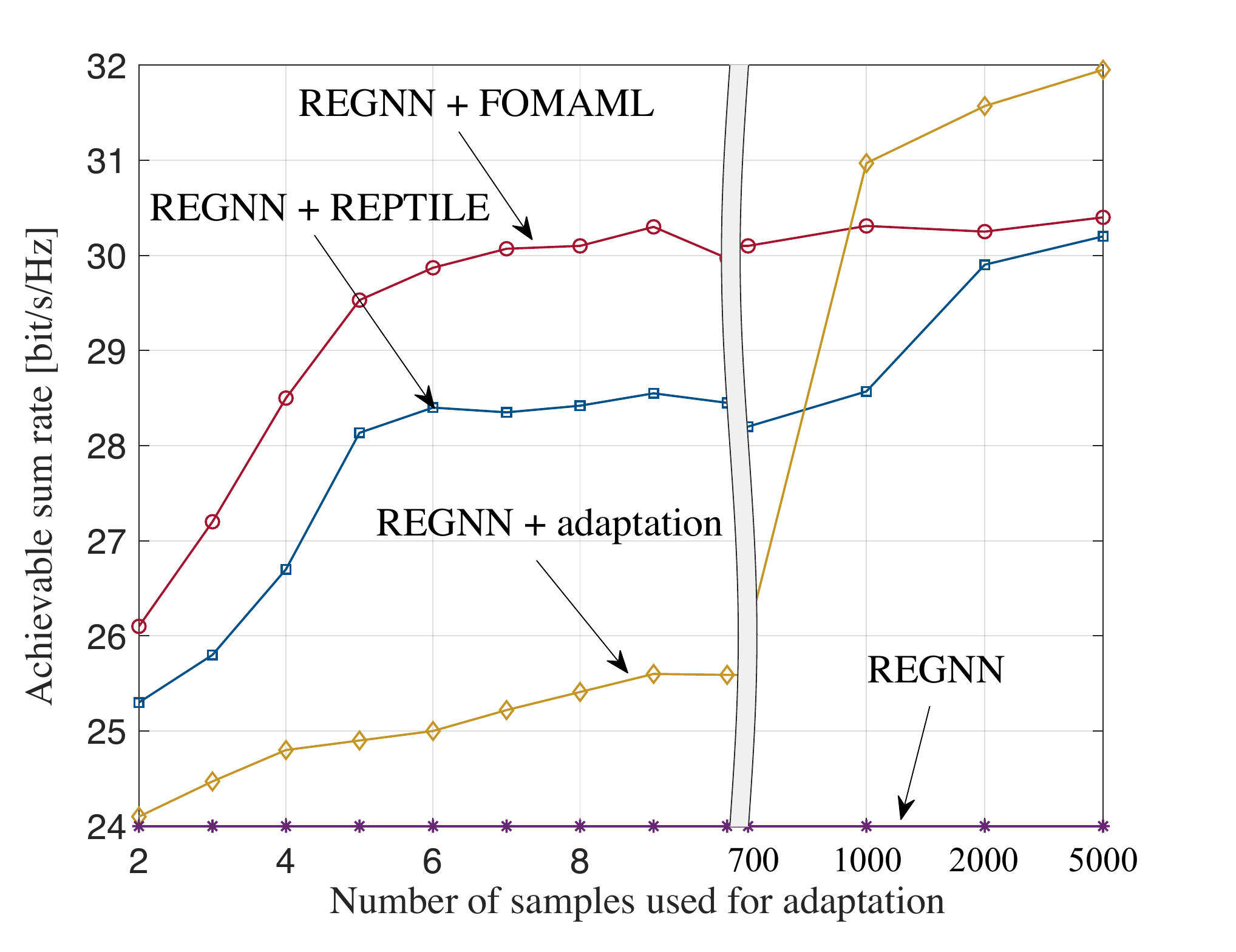}
\caption{Sum rate as a function of the number of samples in the set $\mathcal{D}_{\tau_\text{test}}$ used for adaptation, for a network with dynamic size. The number of training and testing samples in the set $\mathcal{D}_\tau$ are set to $|\mathcal{D}^{\text{tr}}_\tau| = 200$, and $|\mathcal{D}^{\text{te}}_\tau| = 200$, respectively. The results are averaged over $20$ independent runs.}
\label{regnn_dyn}
\end{figure}

The achievable sum rate with respect to the number of samples in the data set $\mathcal{D}_{\tau_\text{test}}$ observed on the new, meta-test, topology at run time is illustrated in Fig.~\ref{regnn_dyn} for a network with dynamic size. Meta-learning, via both FOMAML and REPTILE, is seen to adapt quickly to the the new topology, outperforming conventional REGNN \cite{eisen2020optimal}, both with and without adaptation. REGNN with adaptation carries out training as in \cite{eisen2020optimal} using the meta-training data in the data set $\mathcal{D}^{\text{tr}}_\tau$, $\tau = 1,...,\mathcal{T}^{\text{meta}}$, and then fine tunes the model parameters using the data set $\mathcal{D}_{\tau_\text{test}}$. In contrast, REGNN does not carry out adaptation. FOMAML achieves a sum rate of roughly $30$ bit/s/Hz with only $6$ samples at run time, whilst REGNN with adaptation requires around $900$ samples. As we will further elaborate on below, this significant improvement can be attributed to the variability of the topologies observed across periods in the considered scenario, which makes joint training as in \cite{eisen2020optimal} ineffective (see also \cite{park2020learning}, \cite{simeone2020learning}). When the number of samples for adaptation is sufficiently large, conventional REGNN training as in \cite{eisen2020optimal} outperforms meta-learning, as the initialisation obtained by meta-learning induces a more substantial bias than joint training due to the mismatch in the conditions assumed for the updates on meta-training and meat-testing tasks (i.e., the different number of samples used for meta-training and adaptation).

%In Fig.~\ref{regnn_loc} we present the achievable sum rate as a function of the number of the SGD iterations used for adaptation (cf. \eqref{locM}) when $5$ samples in the data set $\mathcal{D}_{\tau_\text{test}}$ are used for adaptation. REPTILE is seen to require fewer updates to achieve best performance, but FOMAML can achieve a largest rate when more updates are carried out. For all schemes, there is an optimal number of SGD updates, in line with results reported in ...

%\begin{figure}[tbp]
%\centering
%\includegraphics[width=0.9\linewidth]{figures/loc_updates_all.pdf}
%\caption{Sum rate as a function of the number of the SGD iterations used for adaptation, for a network with dynamic size.}
%\label{regnn_loc}
%\end{figure}

\begin{figure}[tbp]
\centering
\includegraphics[width=0.9\linewidth]{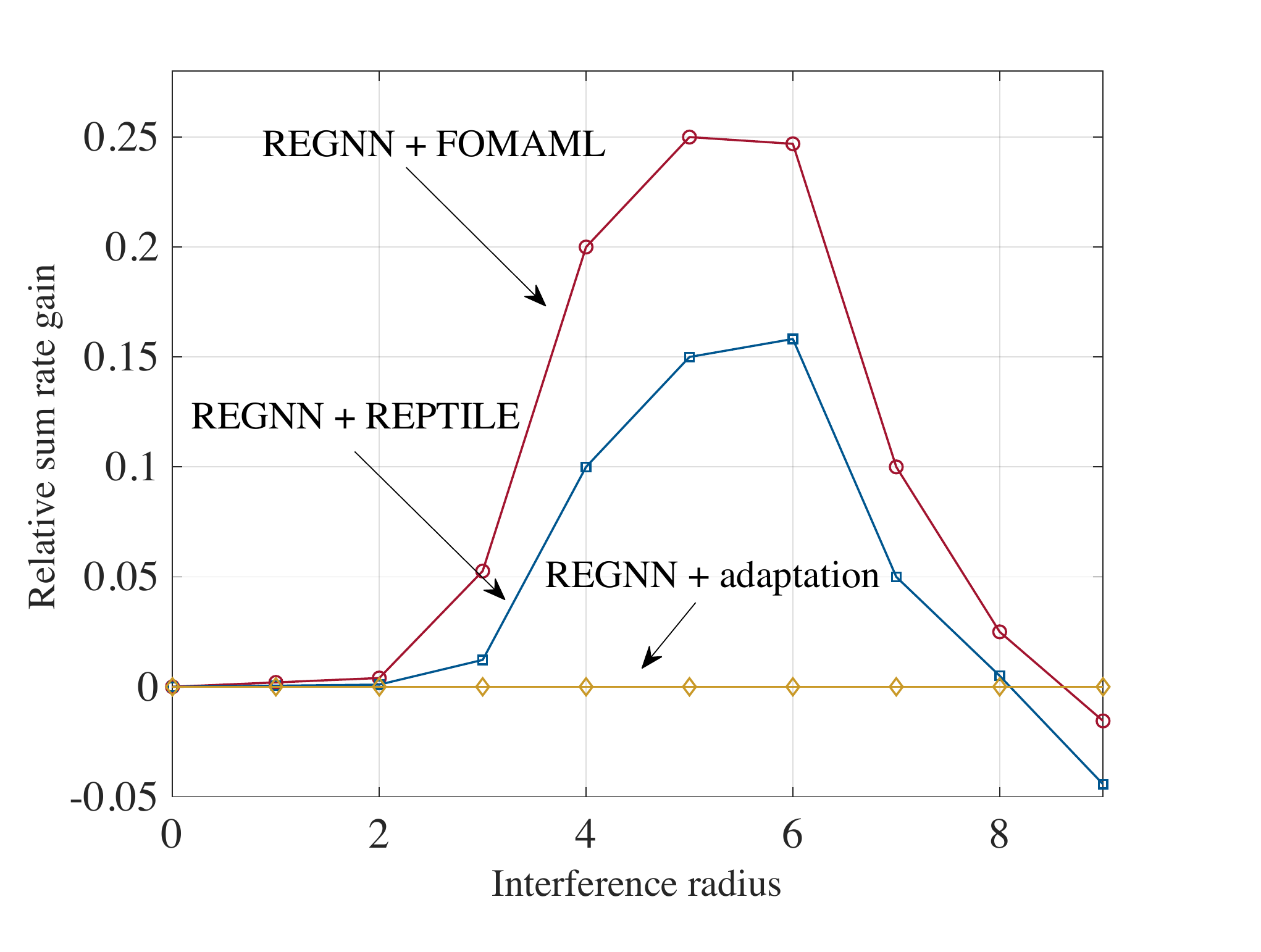}
\caption{Relative rate as a function of the interference radius, for a network of fixed size where $|\mathcal{K}_\tau| = 10$. The number of samples in the set $\mathcal{D}_{\tau_\text{test}}$ used for adaptation is set to $|\mathcal{D}_{\tau_\text{test}}| = 5$. The number of training and testing samples in the set $\mathcal{D}_\tau$ are set to $|\mathcal{D}^{\text{tr}}_\tau| = 200$, and $|\mathcal{D}^{\text{te}}_\tau| = 200$, respectively.}
\label{regnn_rad}
\end{figure}

To understand further how and when meta-learning can improve the efficiency of power allocation, we plot the relative achievable rate as a function of the interference radius in Fig.~\ref{regnn_rad} for a network of fixed size where $\mathcal{K}_\tau = 10$. The relative rate gain is computed as $(C_{\text{ML}}-C_{\text{REGNN+adaptation}})/C_{\text{ML}}$, where $C_{\text{ML}}$ and $C_{\text{REGNN+adaptation}}$ are the sum rates obtained by meta-learning and REGNN with adaptation, respectively. We use $5$ samples in the data set $\mathcal{D}_{\tau_\text{test}}$ for adaptation, and the number of SGD updates is set to $5$, $2$, and $5$, for FOMAML, REPTILE and REGNN with adaptation, respectively. A small radius yields a fully disconnected graph at period $\tau$, while, as the interference radius increases, the graph becomes increasingly connected. At first, this produces a variety of topologies, until only a fully connected graph is obtained for sufficiently large values of the interference radius. Therefore, the distribution of the topologies is maximally diverse at some intermediate value of the interference radius. In line with this observation, meta-learning is seen to profit from task diversity, which prevents meta-overfitting \cite{jose2021information}.

\vspace{-2mm}

\section{Conclusion}\label{sec_con}
In decentralized wireless networks, meta-learning can enable quick adaptation of the power control policy to new network topologies by transferring knowledge from previously observed network configurations. 
This paper investigated the application of meta-learning for adaptation of the power control policy, parameterised by REGNNs, by adopting first-order meta-learning techniques, namely FOMAML and REPTILE. Numerical results have demonstrated that the proposed integration  of  meta-learning and  REGNNs offers significant  improvements  in  terms  of  sample  and  iteration efficiency.

\bibliographystyle{IEEEtran}
\bibliography{litdab.bib}

\end{document}